\documentclass[10pt,journal,twocolumn]{IEEEtran}
\usepackage{graphicx}
\usepackage{dcolumn}
\usepackage{amsmath,amsfonts}
\usepackage{bm}
\usepackage{multirow}
\usepackage{slashbox}
\usepackage{tabularx}
\usepackage{mathtools}
\usepackage{tikz}
\usepackage{amsmath}
\usepackage{amssymb}
\usepackage{xcolor,colortbl}
\definecolor{LightCyan}{rgb}{0.88,1,1}
\definecolor{Gray}{gray}{0.85}
\begin{document}
\title{Neural domain alignment for spoken language recognition based on optimal transport}

\author{Xugang Lu$^{1*}$, Peng Shen$^{1}$, Yu Tsao$^{2}$, Hisashi Kawai$^1$
\thanks{1. National Institute of Information and Communications
Technology, Japan. 
}
\thanks{2. Research Center for Information Technology Innovation, Academic Sinica, Taiwan} }

\IEEEcompsoctitleabstractindextext{%
\begin{abstract}
\small{}
Domain shift poses a significant challenge in cross-domain spoken language recognition (SLR) by reducing its effectiveness. Unsupervised domain adaptation (UDA) algorithms have been explored to address domain shifts in SLR without relying on class labels in the target domain. One successful UDA approach focuses on learning domain-invariant representations to align feature distributions between domains. However, disregarding the class structure during the learning process of domain-invariant representations can result in over-alignment, negatively impacting the classification task. To overcome this limitation, we propose an optimal transport (OT)-based UDA algorithm for a cross-domain SLR, leveraging the distribution geometry structure-aware property of OT. An OT-based discrepancy measure on a joint distribution over feature and label information is considered during domain alignment in OT-based UDA. Our previous study discovered that completely aligning the distributions between the source and target domains can introduce a negative transfer, where classes or irrelevant classes from the source domain map to a different class in the target domain during distribution alignment. This negative transfer degrades the performance of the adaptive model. To mitigate this issue, we introduce coupling-weighted partial optimal transport (POT) within our UDA framework for SLR, where soft weighting on the OT coupling based on transport cost is adaptively set during domain alignment. A cross-domain SLR task was used in the experiments to evaluate the proposed UDA. The results demonstrated that our proposed UDA algorithm significantly improved the performance over existing UDA algorithms in a cross-channel SLR task. 
\end{abstract}
}

\maketitle

\IEEEdisplaynotcompsoctitleabstractindextext

\IEEEpeerreviewmaketitle
\section{Introduction}
\label{sec-I}
Spoken language recognition (SLR) is a widely employed technique in multilingual speech applications to identify language types from acoustic speech signals \cite{Mabin2013,Mabin2007c,Mabin2007b}. The development of robust SLR algorithms for different application scenarios is vital to ensure the widespread availability and effectiveness of these applications \cite{Mabin2013,Mabin2007c,Mabin2007b,Lee2016IS}. The majority of advanced SLR systems are based on deep learning (DL) algorithms, primarily benefiting from immense power DL algorithms for feature exploration using significant training data samples. For example, in most SLR systems, the X-vector-based feature representation (inspired by speaker embedding \cite{Snyder2018}) was adopted for language recognition. The advantage of using an X-vector representation is that the neural network model for X-vector extraction can be efficiently trained using numerous speech samples from various languages. Moreover, to explore robust language information and collect data from various recording scenarios, data augmentation with different noise types and signal-to-noise ratios (SNR) could be easily applied to model training \cite{Snyder2018}. Based on this robust language feature representation, SLR can be designed with a classifier model, either a conventional classifier model (such as the Gaussian mixture model or logistic regression model) or another neural network-based classifier model. A unified modeling strategy can be employed, where feature extraction and classifiers are optimized simultaneously within an end-to-end neural network model for the SLR \cite{RichardsonIS,RichardsonIEEE,Moreno2016,Moreno2014,Diez2015,Fernando2017,Geng2016}. 

Although supervised deep feature and classifier learning, in most studies, have significantly improved the SLR performance, there is a simple assumption that the training (source domain) and test (target domain) datasets share a similar statistical distribution. However, in practice, there are often distribution differences between the training and test data or conditions. These differences can drastically degrade the performance due to distribution shifts or mismatches between the training and test conditions. Although data augmentation can relieve the domain mismatch problem to some degree under a wide range of noisy conditions during feature extraction and classification models, it is impossible to cover all the unknown test domain conditions. Domain adaptation is an effective solution to address these distribution gaps. The purpose of domain adaptation is to align the distributions of the training and test data to match them well. With a large collected labeled testing dataset, it is not difficult to obtain a domain transfer function using supervised learning algorithms. However, in real applications, the label information of the testing dataset is not available. Therefore, this study mainly focuses on a preferable and challenging situation, that is, unsupervised domain adaptation (UDA) for cross-domain SLR, where no label is available in the target domain.  

Several UDA algorithms have been developed to address cross-domain mismatch problems in SLR. For instance, probabilistic linear discriminant analysis (PLDA)-based UDA, originally designed for speaker verification, can be adapted for SLR \cite{Romero2014ODS,Romero2014SLT}. Another approach involves modifying the PLDA framework with feature-based correlation alignment (CORAL) for speaker verification \cite{LeeICASSP2019}. Additionally, feature-distribution adaptation methods that consider the mean and covariance of different domain vectors have also been proposed \cite{Bousquet2019} to solve cross-domain mismatch problems. Neural network-based UDA algorithms have recently become the preferred adaptation framework as they allow for effective adjustment of feature representations and classifiers using gradient-based learning algorithms. In most representative neural-network-based UDA algorithms, an $\mathcal{H}$-divergence or discrepancy metric is designed to measure the feature distribution difference between the source and target domains during domain alignment training; for example, maximum mean discrepancy (MMD) \cite{MMD} and central moment discrepancy (CMD) \cite{CMD}. Moreover, a domain adversarial neural network (DANN) with gradient reversal learning, which was first proposed for image processing \cite{DANN2016}, has been applied to cross-domain SLR tasks \cite{Badr2020}. In a DANN, the domain discriminator and feature generator work as a mini-max game inspired by generative adversarial nets (GANs) \cite{Goodfellow2014}. However, due to the unstable gradient and modal collapse properties of model framework, as demonstrated by multiple studies, it is challenging to train mini-max game-based models. Recently, Wasserstein distance-guided representation learning (WDGRL)-based domain adaptation was proposed \cite{Shen2018}. The advantage of WDGRL is that it can provide a stable gradient in feature learning, even when there is an enormous difference in support of the probability distributions between the source and target domains. Regardless of which optimization algorithm is used, the most crucial aspect is to design a discrepancy metric to measure the distribution difference during domain alignment.    

The majority of UDA algorithms focus on reducing domain shifts by considering the global feature distributions of the source and target domains. However, it is beneficial to align the feature distributions of different domains belonging to the same class, taking into account the local class distribution structures. This approach minimizes the global domain distribution discrepancy and achieves satisfactory performance in the target domain. Although class structures exist in feature distributions (for example, clustering structures), owing to the unavailability of target labels, it is difficult to use them for discrepancy measures based on conventional moment-based discrepancy measurement metrics (such as MMD and CMD). A discrepancy measure utilizing class-wise structure information should perform better when performing domain alignment in UDA. Recently, optimal transport (OT) has been extensively investigated for domain adaptation in machine learning \cite{CourtyIEEE2017}. The initial motivation for OT in machine learning is to determine an optimal transport plan to convert one probability distribution into another with minimal effort \cite{Peyre2018}. Determining optimal transport defines a distance measure between different probability distributions. The OT has an excellent property in that it can consider the geometrical characteristics of distributions when performing optimal transport. Owing to this property, OT is a promising tool for domain adaptation and shape matching in image processing, classification, segmentation, and speech enhancement \cite{CourtyIEEE2017,CourtyNIPS2017, DamodaranECCV2018, Lin2021}. Inspired by the OT-based unsupervised adaptation \cite{CourtyIEEE2017,CourtyNIPS2017, DamodaranECCV2018}, we have proposed an unsupervised neural adaptation framework for cross-domain SLR task \cite{LuICASSP2021}. After using the adaptation model, significant improvements were observed in the cross-domain SLR task. In our previous study, we assumed that the training and test data domains share the same label space. However, in most real SLR tasks, the labeling space of the training dataset is significantly larger than the testing dataset. For example, in the cross-domain SLR task \cite{OLR2020}, the labeling space of the test set is only a subset of that of the training set (10 categories in the training set and only six categories in the test set). Reducing the domain discrepancy by fully matching their distributions may result in a negative effect, that is, a negative transfer by aligning irrelevant classes between different domains. Therefore, in performing OT, training data samples whose categories are not included in the test dataset should not be coupled for distribution transport during adaptation. As an extension and further development of our previous study \cite{LuICASSP2021}, we propose a joint distribution alignment (JDA) model based on partial optimal transport (POT) for the SLR. Our main contributions are summarized as follows: (1) We propose a neural alignment model based on the joint distribution OT for SLR via a neural network modeling framework. Although OT has been extensively applied to image processing and adaptation, we first adapted this concept for cross-domain SLR. In particular, we unified a latent feature projection and softmax-based conditional classifier transform in an end-to-end neural alignment model, where the OT was embedded in a gradient-based learning framework for model parameter optimization. (2) In distribution alignment based on OT, we further proposed that only partial couplings between training and test are allowed in alignment, that is, by setting a threshold of transport cost, training samples with transport costs higher than the threshold will not join in the distribution alignment. The soft weighting of the coupling was adaptively set during distribution transport to reduce the risk of removing possibly matched training samples. Based on this neural partial OT adaptation model, it is supposed that the negative transport effect will be alleviated, thereby improving performance. (3) Based on the proposed UDA, we developed an adaptation system for the SLR and conducted experiments to test its performance. Based on our results, we achieved superior performance compared to several advanced UDA algorithms.

The remainder of this study is organized as follows: Section \ref{sect_frm} presents the theoretical concerns and related work on UDA algorithms. Section \ref{sect_POT} introduces the proposed UDA algorithm and develops the UDA SLR system. In Section \ref{sect_exp}, the experiments conducted are evaluated, and comparisons with several UDA algorithms are presented. Finally, the discussions and conclusions are presented in Section \ref{sect_conclude}. 

\section{Theoretical concern and related work}
\label{sect_frm}
Our proposed neural UDA model for cross-domain SLR is based on the current deep-speaker model, in which the X-vector is used as a language feature representation. X-vector extraction is based on a deep neural network model, for example, a time delay neural network (TDNN) \cite{Snyder2018} and its extensions. Based on the X-vector, a feature projection (as a latent feature space) and a classifier (as a label space) were designed. Based on this deep model, we propose a UDA algorithm that uses an OT-inspired discrepancy metric. Before deriving our proposed model framework and adaptation algorithm, we review the theoretical concerns and related works on UDA in designing the cross-domain adaptation model. 

\subsection{Cross-domain adaptation}
\label{sect-cross}
Given a source domain dataset, $D^s  = \left\{ {\left( {{\bf x}_i^s ,{\bf y}_i^s } \right)} \right\}_{i = 1,..,N}$ and a target domain dataset $D^t  = \left\{ {\left( {{\bf x}_i^t ,{\bf y}_i^t } \right)} \right\}_{i = 1,..,M}$, where the target label information ${\bf y}_i^t $ is not available. $N$ and $M$ are the numbers of samples in the source and target domains, respectively. Owing to domain changes (such as different recording channels in the SLR task), a domain distribution discrepancy exists: $p^s \left( {{\bf x},{\bf y}} \right) \ne p^t \left( {{\bf x},{\bf y}} \right)$. The expected risk or error in selecting a hypothesis transform in the target domain can be formulated as:
\begin{equation}
	\begin{array}{l}
		\begin{aligned}
			R^t (f) &= E_{({\bf x},{\bf y}) \to p^t ({\bf x},{\bf y})} [l(f({\bf x}),{\bf y})] \\ 
			&= \frac{1}{{|\mathcal{C}|}}\sum\limits_{{\bf y} \in \mathcal{C}} {\int {l(f({\bf x}),{\bf y})p^t ({\bf x},{\bf y})} d{\bf x}}  \\ 
			&= \frac{1}{{|\mathcal{C}|}}\sum\limits_{{\bf y} \in \mathcal{C}} {\int {l(f({\bf x}),{\bf y})\frac{{p^t ({\bf x},{\bf y})}}{{p^s ({\bf x},{\bf y})}}} p^s ({\bf x},{\bf y})d{\bf x}}  \\
		\end{aligned}
	\end{array}
\label{eq_risk}
\end{equation}
In this formulation, $f: \mathcal{X} \to \mathcal{Y}$ is a function that transforms the feature $\bf x$ into label $\bf y$, $l(.,.)$ is a risk or loss function that considers the predicted and real labels as inputs, $\mathcal{C}$ is the label space, and $|\mathcal{C}|$ is the size of the label space. From this equation, we note that only when ${\frac{{p^t ({\bf x},{\bf y})}}{{p^s ({\bf x},{\bf y})}}}=1$ can the error in the target domain be maintained the same as that in the source domain. Therefore, the purpose of domain adaptation is to align the probability distributions of the source and target domains ${p^s ({\bf x},{\bf y})}$ and ${p^t ({\bf x},{\bf y})}$. 

For the convenience of the probability distribution alignment between the source and target domains, a suitable latent feature space is required in which the transformed feature from $\bf x$ is present. Correspondingly, in a deep neural network-based domain adaptation of the SLR, the transform function $f(.)$ in Eq. (\ref{eq_risk}) is typically implemented by stacking several layers of a neural network model for suitable feature exploration and classification boundary decisions. For the convenience of analysis, we explicitly regard the transform function as a composition of two functional modules as follows:
\begin{equation}
	y({\bf x}) = f({\bf x};\theta _g ,\theta _\phi  ) = g \circ \phi ({\bf x}),
	\label{eq_comp}
\end{equation}
where mapping functions $\phi:\mathcal{X} \to \mathcal{Z}$ and $g:\mathcal{Z} \to \mathcal{Y}$ are feature extraction and classifier transforms with parameter sets ${\theta _\phi }$ and ${\theta _g }$, respectively, ``$\circ$" is the function composition operator. Accordingly, the latent feature can be obtained as ${\bf z} = \phi \left( {\bf x} \right) $, and the class-wise probability values can be obtained using the classifier module $g\left(  \cdot  \right)$. The process is as follows: $\mathcal{X}\mathop  \to \limits^{\phi } \mathcal{Z}\mathop  \to \limits^{g} \mathcal{Y}$ and adaptation is performed on the two functional modules $g$ and $\phi $. 
   
Based on Eq. (\ref{eq_comp}), the solution in most studies is to discover a latent feature space with a transform ${\bf z} = \phi ({\bf x})$, where the joint distributions can be approximated (or aligned) $p^s ({\bf z},{\bf y}) \approx p^t ({\bf z},{\bf y})$. Based on the Bayesian theory, the joint distribution of features and labels in the transformed space is formulated as:
\begin{equation}
	p^u ({\bf z},{\bf y}) = p^u ({\bf y}|{\bf z})p^u ({\bf z});u = \{ s,t\}
	\label{eq_bayes}
\end{equation}
For a model trained using a source domain dataset, the expected risk or error bound in the target domain can be estimated using the following formula \cite{Chuang2020,LiuBound,Kouw2019}:
\begin{equation}
R^t (f) \le R^s (f) + L_\phi  (p_{\bf z}^s ,p_{\bf z}^t ) + R_{\bf y}  
\label{eqrisk}
\end{equation}
where 
\begin{equation}
	\begin{array}{l}
		\begin{aligned}			
		  R_{\bf y} \mathop  = \limits^\Delta   \min \{ & {E_{D^s } [ {| {p^s ({\bf y}|{\bf z}) - p^t ({\bf y}|{\bf z})} |} ]}, \\ 
			& {E_{D^t } [ {| {p^s ({\bf y}|{\bf z}) - p^t ({\bf y}|{\bf z})} |} ]} \} \\ 
		\end{aligned}
	\end{array}
	\label{eqrisk3}
\end{equation}

In this formulation, $R^t (f)$ is risk or generalization error in target domain and $R^s (f)$ is the risk in source domain (as $E_{({\bf x},{\bf y}) \to p^s ({\bf x},{\bf y})} [l(f({\bf x})),{\bf y}]$). $L_\phi (p_{\bf z}^s ,p_{\bf z}^t )$ is a divergence measure between the marginal distributions of source and target domains ($p_{\bf z}^s$ and $p_{\bf z}^t$ are short notes for $p^s ({\bf z})$ and $p^t ({\bf z})$ respectively). Based on this formulation, we note that the expected risk in the target domain is closely related to three terms: risk in the source domain, feature distribution discrepancy between the two domains, and distances of labeling functions between the source and target domains. The cross-domain adaptation objective is lowering the upper bound of the target risk, or generalization error bound $R^t (f)$. In most studies, the first and second terms of Eq. (\ref{eqrisk}) are considered, whereas the third term is small and ignored. Based on this consideration, most studies aim to explore domain invariant representations while maintaining small risks in the source domain, which are introduced in the following. 
\subsection{Domain invariant representation learning}
The concept of domain-invariant representation learning is to discover a latent feature space in which the feature projections of samples from the source and target domains maintain similar probability distributions. Hence, the classifier trained with the source domain dataset can be directly used for the target domain data. There are two model frameworks for domain-invariant feature learning: statistic moment matching with direct minimization of the feature distribution discrepancy between the source and target domains, and adversarial learning by learning the feature generator and domain discriminator in a min-max game.
\subsubsection{Minimizing distribution discrepancy based learning}
In this adaptation framework, two types of losses are involved in the objective function for model parameter optimization: Classification loss in the source domain and feature distribution discrepancy loss between the source and target domains. Specifically, the model parameters were optimized by minimizing the following objective function:
\begin{equation}
L(\theta _\phi  ,\theta _g ) = \sum\limits_i {L_{{\rm CE}}^s ({\bf y}_i^s ,{\bf \hat y}_i^s ) + \lambda L_{\phi} (p_{\bf z}^s ,p_{\bf z}^t )}   
\label{eq_DMMD1}
\end{equation}
\begin{equation}
\theta _\phi ^* ,\theta _g^* \mathop  = \limits^\Delta  \arg \mathop {\min }\limits_{\theta _\phi  ,\theta _g } L(\theta _\phi  ,\theta _g ),
\label{eq_DMMD2}
\end{equation}	
where $L_{\phi} (p_{\bf z}^s ,p_{\bf z}^t )$ is the discrepancy measure of the latent feature distributions between the source and target domains. In addition, $L_{{\rm CE}}^s ({\bf y}_i^s,{\bf \hat y}_i^s)$ is the cross-entropy loss in the source domain, and ${\bf \hat y}_i^s $ is the predicted label of a sample (indexed by $i$) from the source domain. The classification risk in the target domain is reduced by minimizing the distribution discrepancy of the latent features between the source and target domains while maintaining a low classification risk in the source domain. Many adaptation models that can be applied to SLR belong to this category, such as those used in MMD-based domain adaptation \cite{PanMMD2009,LongMMD2013} and CMD-based domain adaptation \cite{MMD}.  
\subsubsection{Domain adversarial learning}
\label{sub_DAL}
In contrast to the direct minimization of feature distribution discrepancy loss-based learning, a domain discriminator is designed to judge whether the data samples are from the source or target domain, while the neural feature generator attempts to explore domain-invariant features to confuse the domain discriminator \cite{LiuBound, DANN2016}. Several studies were inspired by \cite{DANN2016} for neural-network-based domain adversarial learning, where the objective function is defined as:
\begin{equation}
L(\theta _\phi  ,\theta _g ,\theta _{\varphi} ) = \sum\limits_i {L_{{\rm CE}}^s ({\bf y}_i^s ,{\bf \hat y}_i^s ) - \lambda {\sum\limits_{\mathclap{{\bf z}_i  \in \{ D^s  \cup D^t \}} }}  {L_{\varphi} (\varphi({\bf z}_i ),d_i )} }  \\ 
\label{eq_adv}
\end{equation}	
	
\begin{align}
\theta _\phi ^* ,\theta _g^* & \mathop   =   \limits^\Delta  \arg \mathop {\min }\limits_{\theta _\phi  ,\theta _g } L(\theta _\phi  ,\theta _g ,\theta _{\varphi} ) \\ 
\theta _{\varphi}^* & \mathop  =   \limits^\Delta  \arg \mathop {\max }\limits_{\theta _{\varphi} } L(\theta _\phi  ,\theta _g ,\theta _{\varphi} )  
\label{eq_minmax}
\end{align}
In this neural network-based adversarial learning, $L_{\varphi} (\varphi({\bf z}_i ),d_i )$ is the domain classification loss with domain label $d_i=1$ for ${\bf z}_i  \in D^s $, and $d_i=0$ for ${\bf z}_i \in D^t $. The neural domain discriminator $\varphi(.)$ is a neural network-based model with the parameter set denoted by $\theta_{\varphi}$. From Eqs. (9) and (10), we can note that the optimization of Eq. (\ref{eq_adv}) is a mini-max game widely used in GAN-based adversarial learning \cite{Goodfellow2014}. Based on this mini-max game, the distributions of latent features in the source and target domains should be aligned while maintaining the classification accuracy in the source domain. 
\subsection{Optimal transport}
In the domain-adaptation algorithms reviewed above, although the feature distributions can be matched well by domain-invariant feature learning, the discriminative power of the features may be degraded \cite{Shen2018, CourtyNIPS2017}. Moreover, although statistical moment matching (for example, MMD- or CMD-based discrepancy measure) can perform well, as used in most domain-adaptation algorithms, the complex geometric structure of the distributions is not explicitly considered. The complex geometric structure of a feature distribution (such as multimodal cluster distributions) always encodes class-wise feature discriminative information. A discrepancy measure that considers this distribution geometry structure will aid in improving the performance of the adaptation model in UDA. Recently, optimal transport (OT)-based theory has been widely applied in machine learning and domain adaptation \cite{CourtyIEEE2017,CourtyNIPS2017}. The basic concept of the OT is defined as the minimum amount of total mass required to move when transforming one probability distribution into another. The Wasserstein distance is a discrepancy measure between two probability distributions inspired by the OT. The Wasserstein distance or OT distance-based discrepancy metric performed satisfactorily in numerous machine learning tasks because class distribution structures are encoded well based on measurements \cite{Shen2018, WGAN2017}. Based on the dual representation by the Kantorovich-Rubinstein theorem \cite{VillanoBook}, the definition of the Wasserstein distance between the source and target distributions is further cast as: 
\begin{equation}
	L_h (p_{\bf z}^s ,p_{\bf z}^t ) = \mathop {\sup }\limits_{h \in \mathcal{L}} E_{{\bf z} \to p_{\bf z}^s } [h({\bf z}))] - E_{{\bf z} \to p_{\bf z}^t } [h({\bf z})]
	\label{eq_KR}
\end{equation}     
where $h(.)$ is a Lipschitz function that maps the latent feature vector (${\bf z} = \phi ({\bf x})$) to a real value and $\mathcal{L}$ is a set of 1-Lipschitz functions. In neural network modeling, an empirical approximation of the expectation in Eq. (\ref{eq_KR}) is adopted as follows:
\begin{equation}
	L_{wd} ({\bf z}^s ,{\bf z}^t ) \mathop  = \frac{1}{{|D^s| }}\sum\limits_{\mathclap{{\bf z}_i  \in D^s }} {h({\bf z}_i )}  - \frac{1}{{| D^t| }}\sum\limits_{\mathclap{{\bf z}_i  \in D^t }} {h({\bf z}_i )}, 	
	\label{eqwdapp}
\end{equation}
where $| D^s|$ and $|D^t|$ are the number of samples in the source and target domains, respectively. In addition, the 1-Lipschitz constraint of the neural network mapping function is implemented as a gradient penalty \cite{Shen2018,ImprovedGAN}:
\begin{equation}
	L_{grad} ({\bf \tilde z})\mathop  = \limits^\Delta  \left( {\left\| {\nabla _{{\bf \tilde z}} h({\bf \tilde z})} \right\|_2  - 1} \right)^2, 
	\label{eq_gradpen}
\end{equation}
where $\nabla$ is the gradient operator and $\bf{\tilde z}$ is a random point along the straight line between the source and target pairs. Based on the definitions given in Eqs. (\ref{eqwdapp}) and (\ref{eq_gradpen}), adversarial learning in the min-max game is formulated as follows:
\begin{equation}
	L_h \mathop  = \limits^\Delta  \mathop {\min }\limits_{\theta _\phi  } \mathop {\max }\limits_{\theta _h } L_{wd} ({\bf z}^s ,{\bf z}^t ) - \eta L_{grad} ({\bf \tilde z}),	
	\label{eqwdgrl}
\end{equation}
where $\eta$ is the penalty coefficient. The neural parameter set of the neural-domain critical function $h(.)$ is denoted in this formula as $\theta _h$. Because transform $\phi(.)$ is involved in the latent feature extraction, the parameter set $\theta _\phi$ is also involved in this adversarial learning.    

When verifying the upper bound of the adaptation model in Eq. (\ref{eqrisk}), we note that in most UDA methods explained above, the classifier for the label space in adaptation is not involved in optimization, that is, only the first two error terms are considered, while the third term is ignored. To incorporate label information in domain adaptation, a joint distribution OT (JDOT) that includes label information in transport cost estimation has been proposed \cite{CourtyIEEE2017,CourtyNIPS2017}. Hence, domain alignment based on JDOT can match distributions by considering both feature and label information between domains. Inspired by this JDOT \cite{CourtyIEEE2017,CourtyNIPS2017,DamodaranECCV2018}, we have proposed a neural OT-based adaptation model for cross-domain SLR \cite{LuICASSP2021}. In contrast to most JDOT-based UDA algorithms, we built our adaptation model on a pre-trained X-vector for SLR, where the latent feature projection and classifier transforms were unified in an end-to-end optimization framework. Based on the adaptation model, we obtained a promising performance in a cross-domain SLR task \cite{LuICASSP2021}. 

\section{Proposed adaptation model for SLR}
\label{sect_POT}
In our previous neural OT-based adaptation model for SLR \cite{LuICASSP2021}, we discovered a negative transfer during transport optimization, that is, a class or an irrelevant class in the source domain is mapped to a different class in the target domain. This negative transfer significantly degrades performance because the adaptation process accumulates incorrect information for the distribution alignment between domains. Two main reasons may induce this negative transfer, one is from the ``harder" examples, which are challenging or ambiguous inaccurate label prediction, and the other is an inappropriate assumption about the label space for the source and target domains in OT. In our previous study, we assumed that the source and target datasets share the same label space size. However, in real applications, for example, in one test set of our SLR task, the label space of the target data is only a subset of the source data. It is unsuitable for aligning distributions with non-consistent classes from the source and target domains. This study proposes a weighted optimal transport algorithm that allows only partial transport for domain alignment. As an extension of our previous work, we briefly introduce our previous neural OT-based adaptation framework for SLR and then derive a new UDA algorithm. 

\subsection{Neural domain adaptation model for SLR}
\begin{figure}[tb]
	\centering
	\includegraphics[width=7cm, height=5cm]{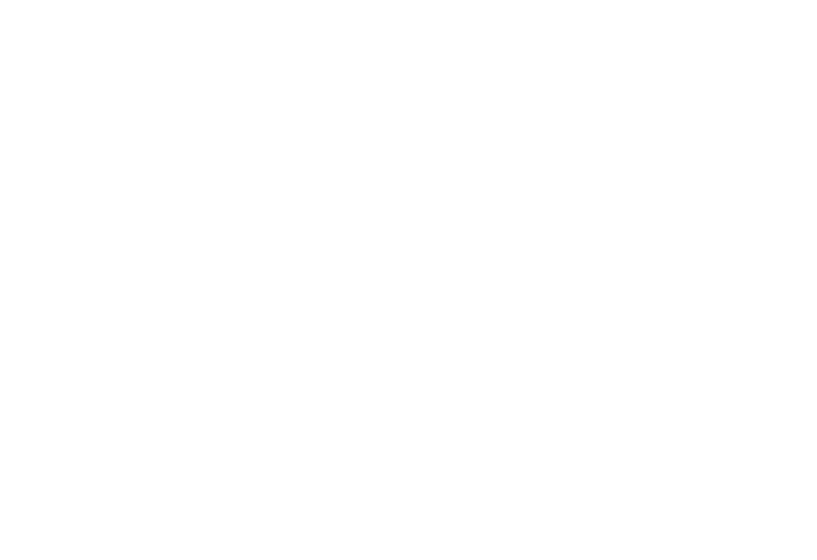}
	\caption{Domain adaptation model framework for SLR based on X-vector extraction and classifier models}
	\label{figConv}
\end{figure}

From Eq. (\ref{eq_bayes}), the joint distribution adaptation can be implemented as a feature marginal distribution alignment and a classifier conditional distribution alignment as follows:
\begin{align}
p^s ({\bf z})  &\approx p^t ({\bf z}) \label{eq:1} \\
p^s ({\bf y}|{\bf z})  &\approx p^t ({\bf y}|{\bf z}) \label{eq:2} 		
\end{align}
In Eq. (\ref{eq:1}), an approximation was used to align the latent feature distribution. The approximations of Eq. (\ref{eq:2}) is used to align the label conditional distribution. Based on this explanation, the proposed SLR adaptation model framework is illustrated in Fig. \ref{figConv}.
In this figure, the encoder block is fixed throughout the training of the adaptation model and is utilized for X-vector extraction. The red dashed-line box represents the feature projection module, where the X-vector is projected into a low-dimensional space for discriminative language feature extraction. The conventional framework comprises linear discriminant analysis (LDA) and vector normalization. A dense-layer neural network model with vector length normalization (L-norm) is utilized in our framework. Based on the normalized feature, a classifier for language category decisions was designed as another dense-layer neural network model with softmax activation (blue dashed box in Fig. \ref{figConv}). In the adaptation, only feature projection and classifier layers with a relatively small number of parameters were adapted; meanwhile, the parameters of the X-vector extraction network were fixed after the encoder model was trained. In our adaptation model framework, both training and testing share the same neural network transforms, and the adaptation process minimizes the distribution discrepancy of the latent features and classifiers between the source and target datasets, as illustrated in Fig. \ref{figConv}. The measure of distribution discrepancy was designed based on the OT distance, which is introduced in the following section.

\subsection{Joint distribution optimal transport distance}
Based on the definition of the Kantorovich optimal transport \cite{VillanoBook}, the discrete OT distance for two probability distributions is defined as follows:
\begin{equation}
L_{{\rm OT}} (p^s ,p^t )\mathop  = \limits^\Delta  \mathop {\min }\limits_{\gamma  \in \prod {(p^s ,p^t )} } \sum\limits_{i,j} {L({\bf v}_i^s ,{\bf v}_j^t )\gamma ({\bf v}_i^s ,{\bf v}_j^t )},
\label{eq_ot}
\end{equation}
where $p^s$ and $p^t$ are probability distributions of source and target domains, respectively. ${\gamma ({\bf v}_i^s, {\bf v}_j^t )}$ is the transport plan (or coupling) between the two distributions sampled from the joint probability space $\prod {(p^s, p^t )}$. In addition, ${L({\bf v}_i^s, {\bf v}_j^t )}$ is a pair-wise cost (or ground cost) between two examples ${{\bf v}_i^s }$ and ${{\bf v}_j^t }$ which are sampled from the two probability distributions $ p^s$ and $p^t$, respectively. In most OT-based adaptation models, this OT distance-based loss is applied for locating a marginal latent feature space where the sample pairs are from $p^s ({\bf z})$ and $p^t ({\bf z})$ (in this case, ${\bf v}_i^s={\bf z}_i^s$, ${\bf v}_j^t={\bf z}_j^t$). Furthermore, it is later modified for joint distribution adaptation framework, that is, both latent feature and label distributions are adopted in image classification \cite{CourtyIEEE2017, CourtyNIPS2017, DamodaranECCV2018}, where the sample pairs are from joint probability distributions $p^s ({\bf z},{\bf y})$ and $p^t ({\bf z},{\bf y})$ (in this case, the input as tuples, ${\bf v}_i^s=\{({\bf z}_i^s ,{\bf y}_i^s)\}$, ${\bf v}_i^t=\{({\bf z}_i^t ,{\bf y}_i^t)\}$), respectively. Correspondingly, the ground cost for joint distributions is defined as:
\begin{equation}
	L({\bf v}_i^s ,{\bf v}_j^t ) \mathop  = \limits^\Delta d_{\bf z} ({\bf z}_i^s ,{\bf z}_j^t ) + \alpha d_{\bf y} ({\bf y}_i^s ,{\bf y}_j^t ), 
	\label{eq_jdist}
\end{equation}
where $d_{\bf z} (.)$ and $d_{\bf y} (.)$ are the distance functions (for example, Euclidean distance) for the latent feature and label, respectively. In Eq. (\ref{eq_jdist}), $\alpha$ is a weighting coefficient that controls the relative importance of the label information for adaptation in cost estimation during OT optimization. From this definition, we note that in this JDOT, both feature and label information are involved in the transport cost estimation. Because the target label is unknown in UDA, the predicted target label is used instead, that is, $d_{\bf y} ({\bf y}_i^s ,{\bf y}_j^t) $ is replaced with $d_{\bf y} ({\bf y}_i^s ,{\bf \hat y}_j^t )$, where ${\bf \hat y}_j^t  = g \circ \phi ({\bf x}_j^t )$. We applied this idea to the SLR and obtained a significant performance improvement \cite{LuICASSP2021}. In our study and implementation, we discovered negative transport, which degrades the performance, that is, misspecified couplings or mappings of samples from different classes between the two domains. It is preferable that samples sharing the same labels between the source and target domains be coupled for transport, that is, partial optimal transport. A conventional solution is to pad zeros in the coupling matrix on misspecified mapping locations during the OT. However, in real applications, the label information for the target domain is unknown, and it is unknown where the misspecified mappings are induced, making it difficult to determine which coupling pairs should be padded with zeros. Therefore, weight transport couplings are proposed to decrease negative transport.
\subsection{Partial optimal transport}
\begin{figure}[tb]
	\centering
	\includegraphics[width=6cm, height=4cm]{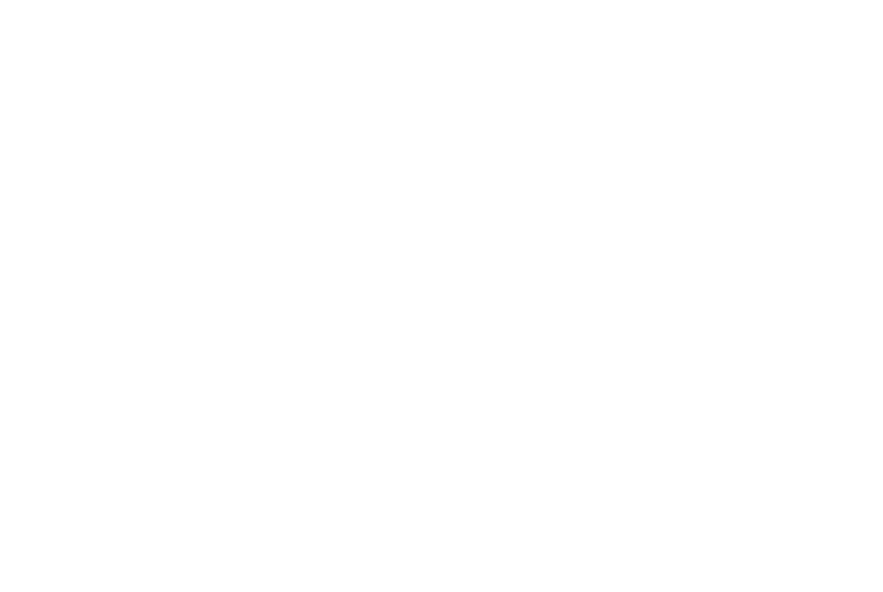}
	\caption{Partial coupling in optimal transport (admissible transport paths) between training (source) and testing (target) samples (refer to text for details).}
	\label{figPcoupling}
\end{figure}

In Eq. (\ref{eq_ot}), the OT distance is defined based on the assumption that the total probability masses of $p^s$ and $p^t$ are equal and that all probability masses from the source $p^s$ are transported to the target $p^t$. In real situations, when the label spaces of the source and target domains are different, it is unsuitable to transport all probability masses from the source to the target. Therefore, the coupling function ${\gamma  \in \prod {\left( {p^s ,p^t } \right)} }$ should be constrained, that is, partial OT (POT), by maintaining only the admissible couplings between the probability masses induced by shared label classes in both the source and target domains. The POT problem was discussed in \cite{Figalli2010}, and partial Wasserstein and Gromov-Wasserstein problems were addressed in \cite{Chapel2020}. In these studies, the probability masses of the source and target domains were intrinsically different and should not be fully coupled in mass transportation. Based on this consideration, Eq. (\ref{eq_ot}) is changed to:
\begin{equation}
	L_{{\rm POT}} (p^s ,p^t )\mathop  = \limits^\Delta  \mathop {\min }\limits_{\gamma  \in \prod ^{{\rm par}} (p^s ,p^t )} \sum\limits_{i,j} {L({\bf v}_i^s ,{\bf v}_j^t )\gamma ({\bf v}_i^s ,{\bf v}_j^t )}
	\label{eq_pot}
\end{equation}
Notably, in this POT-based distance, the couplings ${\gamma \in \prod ^{{\rm par}} (p^s ,p^t )}$ are only partially admissible between the source and target domains, which differ from those used in Eq. (\ref{eq_ot}). For convenience, the partial transport coupling between the training and testing samples is illustrated in Fig. \ref{figPcoupling}.
In this figure, the vertical samples are from the source domain with labeling sets $\{ {\rm C0} ,{\rm C1} ,{\rm C2} \}$ (different symbols represent different classes, as illustrated in Fig. \ref{figPcoupling}), the horizontal samples are from target domain with labeling set $\{ {\rm C1} ,{\rm C2} \}$. Admissible couplings are marked as $1$, otherwise they are marked as $0$. Although this POT can easily eliminate the negative transfer effect by ignoring the non-admissible couplings, in real situations, the target labels are unknown; instead, we can set a threshold of the transport cost to obtain the admissible couplings as:
\begin{equation}
	w_{i,j}  = \left\{ \begin{array}{l}
		1,L({\bf v}_i^s ,{\bf v}_j^t ) \le \tau  \\
		0,L({\bf v}_i^s ,{\bf v}_j^t ) > \tau  \\
	\end{array} \right.
\end{equation}
where $\tau $ is the transport cost threshold and ${\bf v}_i^s$ and ${\bf v}_j^t$ are the two samples from the source and target domains indexed by $i$ and $j$, respectively. In this equation, if the transport cost is less than $\tau $, coupling between the two samples is allowed; otherwise, coupling is discarded. However, this difficult control of couplings may pose a risk in discarding admissible or accepting non-admissible couplings. Therefore, soft weighting was applied to the coupling, which is defined as follows:
\begin{equation}
	\tilde w_{i,j}  = \sigma \left( { - \beta *\left( {L({\bf v}_i^s ,{\bf v}_j^t ) - \tau} \right)} \right),
	\label{eq_softw}
\end{equation}
where $\sigma(.)$ is a sigmoid function and $\beta$ is a scaling parameter. Finally, the solution of the POT is changed to:
\begin{equation}
	L_{{\rm POT}} (p^s ,p^t )\mathop  = \limits^\Delta  \mathop {\min }\limits_{\gamma  \in \prod (p^s ,p^t )} \sum\limits_{i,j} {L({\bf v}_i^s ,{\bf v}_j^t )\gamma ({\bf v}_i^s ,{\bf v}_j^t )\tilde w_{i,j} }
	\label{eq_potfor}
\end{equation}
In this formulation, although the coupling ${\gamma \in \prod (p^s, p^t )}$ is the same as that used in the conventional OT, the soft weighting defined in Eq. (\ref{eq_softw}), which can be regarded as the uncertainty of coupling, is explicitly added to fulfill the function of POT. The POT-based distance in the subsequent section is integrated into neural network learning for cross-domain SLR.

 \subsection{Neural domain alignment based on POT}
 \label{subsect_NNOT}
\begin{figure}[tb]
	\centering
	\includegraphics[width=8cm, height=4cm]{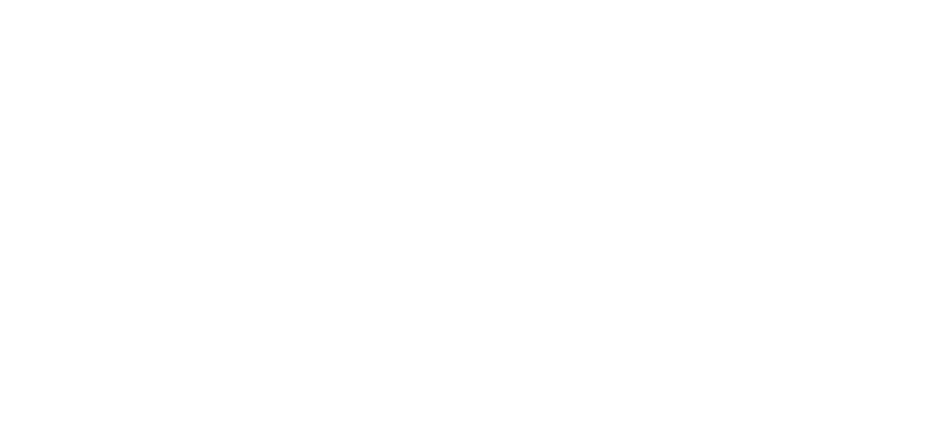}
	\caption{Optimization of neural alignment model based on OT.}
	\label{figcoupling}
\end{figure}

Based on the definition of POT in the neural adaptation model, the total loss, including the domain alignment loss and classification loss in the source domain, is formulated as:
\begin{equation}
		L_{\rm T} (\gamma {\rm ,}\theta _g ,\theta _{\phi}  ) = \sum\limits_i {L_{{\rm CE}}^s ({\bf y}_i^s ,{\bf \hat y}_i^s )}  + \lambda L_{{\rm POT}} (p^s ,p^t ) \\ 
	\label{eq_TLoss}
\end{equation}
In this loss function, the classification loss in the source domain is defined as multiclass cross-entropy:
\begin{equation}
L_{{\rm CE}}^s \left( {{\bf y}_i^s ,{\bf \hat y}_i^s } \right) =  - \sum\limits_{j = 1}^C {y_{i,j}^s \log \hat y_{i,j}^s },
\end{equation}
where ${{\bf \hat y}_i^s }$ is the estimated label for the sample in the source domain and ${\bf \hat y}_i^s = f\left( {{\bf x}_i^s } \right) = g \circ \phi \left( {{\bf x}_i^s } \right)$ ($i$ is the sample index). The model parameters can be obtained by minimizing the loss defined in Eq. (\ref{eq_TLoss}) as:
\begin{equation}
		\gamma _{{\rm POT}}^* ,\theta _g^* ,\theta _{\phi}^* \mathop  = \limits^\Delta  \mathop {\arg \min }\limits_{\gamma {\rm ,}\theta _g ,\theta _h } L_{\rm T} (\gamma {\rm ,}\theta _g ,\theta _{\phi} )
	\label{eq_TLossmini}
\end{equation}
The parameter optimization in Eq. (\ref{eq_TLossmini}) is a two-step process in which the OT distance itself is minimized by first obtaining the optimal coupling as follows: 
\begin{equation}	
	\gamma _{_{{\rm POT}} }^* \mathop  = \limits^\Delta  \mathop {\arg \min }\limits_{\gamma  \in \prod {(p^s ,p^t )} } \sum\limits_{i,j} {L({\bf v}_i^s ,{\bf v}_j^t )\gamma ({\bf v}_i^s ,{\bf v}_j^t )\tilde w_{i,j} }  		
	\label{eq_gamma}
\end{equation}
Based on the optimal coupling matrix $\gamma _{_{{\rm POT}} }^* $, the OT distance is estimated based on:
\begin{equation}
	L_{{\rm POT}} (p^s ,p^t ) = \sum\limits_{i,j} {L({\bf v}_i^s ,{\bf v}_j^t )\gamma _{_{{\rm POT}} }^* (i,j)}  
\end{equation}
where $\gamma _{_{{\rm POT}} }^* (i,j)$ is the $(i,j)$-th element of $\gamma _{_{{\rm POT}} }^* $. 
The total loss minimization in Eq. (\ref{eq_TLoss}) is processed based on the POT loss. The OT was embedded in an end-to-end neural network framework for model optimization in implementation, as illustrated in Fig. \ref{figcoupling}.
In this model framework, three parameter sets were optimized ( Eq. (\ref{eq_TLossmini})), that is, the optimal transport plan matrix $\gamma$ as estimated based on Eq. (\ref{eq_gamma}), the feature transform-related parameters ${\theta _{\phi} }$, and the classifier-related parameters ${\theta _g }$ as defined in Eq. (\ref{eq_comp}). As illustrated in Fig. \ref{figcoupling}, estimating the optimal transport plan matrix requires feature extraction for both source and target domains (to calculate the cost matrix); meanwhile, the optimal feature extraction and prediction depend on the optimal transport plan. The optimization can be incorporated into a bi-level optimization learning framework with a mini-batch sampling of the source and target samples, that is, the OT algorithm, which is used for domain distribution alignment, is embedded in the neural network model and, together with the feature representation learning, is optimized in an end-to-end neural model framework. 

In semi-supervised learning, regularization with entropy minimization is supposed to improve the prediction performance in the target domain \cite{SSLEntropy}. Based on this idea, the data cluster structure in the target domain, which represents class discriminative information, is considered with entropy minimization regularization, defined as:
\begin{equation}
	H_t  = \sum\limits_{{\bf x}_i  \in D^t } {H(g \circ \phi ({\bf x}_i ))}
	\label{eq_entropytarget}
\end{equation} 
where $H(.)$ is the information entropy function. Adding this constraint can drive the model to identify classification boundaries along low-density regions. Because entropy regularization is a universal constraint in any adaptation model, in our study, we added this constraint to our proposed adaptation model and all comparison models. 

\section{Experiments and evaluations}
\label{sect_exp}
Experiments were conducted to evaluate the proposed adaptation model and to compare it with several advanced UDA algorithms for cross-domain SLR tasks. The experimental data corpus was obtained from the Oriental Language Recognition (OLR) Challenge \cite{OLR2020,WD2019,WD2020}. The training dataset was obtained from multilingual OLR and THCHS30 databases, designed for the challenge in previous years (2016, 2017) \cite{WD2020,OLR2020}. As a source domain dataset, there are approximately 110k utterances (more than 100 h) from 10 languages (Mandarin, Cantonese, Indonesian, Japanese, Russian, Korean, Vietnamese, Kazakh, Tibetan, and Uyghur). For the target datasets, two test sets on different tasks were performed: one set was for the short utterance SLR task (corresponding to task 1 in the OLR 2019 challenge), and the other was a cross-channel SLR test (corresponding to task 2 in the OLR 2019 challenge). Concerning the labeling space of the target datasets, the Task 1 test set includes utterances for the same 10 languages as the training set (1.8k utterances for each); however, the utterance duration is short (1s). Task 2 included only six languages (a subset of the labeling space of the training set), also with 1.8k utterances for each. The utterances were recorded in wild environments, quite different from those in the training dataset. Task 2 comprised one development set and one test set. For a slightly different scenario, each was used as an independent target dataset for the UDA model. Four UDA algorithms for SLR are compared. Two of them are based on the directed minimizing distribution discrepancy method, as defined in Eq. (\ref{eq_DMMD1}) and (\ref{eq_DMMD2}), and the discrepancy losses were defined based on the MMD and CMD, respectively. The other two models were based on domain-adversarial learning, as defined in Eqs. (\ref{eq_adv}), (9) and (10): One is with the domain discriminator designed the same as in DANN \cite{DANN2016}, and the other is based on WDGRL \cite{Shen2018}. The basic concepts behind these algorithms are described in Section II. 

In this study, two evaluation metrics were adopted by considering the missing and false alarm probabilities for target and non-target language pairs to measure the performance of the adaptation models. The equal error rate (EER) is when the false acceptance rate (FAR) and false rejection rate (FRR) are equal on the Receiver Operating Characteristic (ROC) curve. The average performance cost (Cavg) is the average of the pairwise loss for target/non-target language pairs while considering the missing and false alarm probabilities \cite{WD2020}. 

\subsection{Language embedding based on X-vector}
The X-vector was extracted using a neural network-based language-embedding model. Similar to speaker embedding \cite{Snyder2018}, the neural network model is an extended time-delay neural network (TDNN) architecture with a bottleneck layer for the X-vector extraction, as implemented in the baseline model \cite{OLR2020}. The extended TDNN is trained based on the cross-entropy criterion for language classification using the OLR training dataset. Data augmentation techniques were applied to improve robustness. The input features for training the language embedding model were 30-dimensional MFCCs extracted from 40 Mel filter bands. The Mel filter band features were extracted using a frame length of 25 ms and a frame shift of 10 ms. Energy-based voice activity detection (VAD) removes silent background regions during the language feature extraction. The final extracted X-vectors had 512 dimensions. Further details on the model architecture and feature extraction are provided in \cite{Snyder2018,OLR2020, WD2020}.
\subsection{Backend modeling}
\label{sectbackend}
Although the X-vector extracted from the language-embedding model is intended to encode language-discriminative information, it also encodes other acoustic factors. Before it is input to a classifier for classification, a language-discriminative analysis with LDA is applied, that is, an LDA-based dimension reduction is adopted to extract language-discriminative information (in this study, the 512-dimension X-vectors are transformed to 200-dimension vectors by the LDA). In most speaker-verification tasks, a PLDA-based generative model is among the most efficient classifiers because the sample size for speaker enrollment is small. However, in the SLR task, multiple mixtures of the logistic regression (LR) model, is regarded among the most robust classifiers because there are numerous samples for each language in the model. Therefore, in our baseline system, the backend system is based on LR after feature normalization from the LDA transform. Moreover, for convenience in implementing the proposed neural OT-based and other types of neural network model-based UDA algorithms, we designed a unified neural backend model, where the classifier was composed of a dense layer with a softmax activation function, as illustrated in Fig. \ref{figConv}. From this figure, the advantage of this backend model-design is that dimension reduction and length normalization are integrated into the optimization process. Although the X-vector extraction network can also be integrated in adaptation learning, to avoid overfitting owing to numerous model parameters, the network parameters for the X-vector extraction model were fixed in this study. The adaptation process was optimized using the Adam algorithm with an initial learning rate of $0.001$ \cite{Adam} and a 128 mini-batch size. Early stopping was used to determine the best model by evaluating its performance on a randomly selected validation set.  
\subsection{Results}
\subsubsection{Baseline performance}
We first evaluate the performances of the two baseline models introduced in Section \ref{sectbackend}: an LR-based backend and a neural network backend models.  
\begin{table}[tb]
	\centering
	\caption{Baseline performances for models with neural back-end and LR back-end classifiers.}
	\begin{tabular}{|c||c||c|}
		\hline
		Test sets&EER(\%) &Cavg\\
		\hline
		\hline
		T1\_test  &\textbf{7.65} (8.41) &\textbf{0.074} (0.082) \\
		\hline
		T2\_dev  &21.13 (\textbf{20.48}) &0.213 (\textbf{0.209})\\
		\hline
		T2\_test  &\textbf{23.89} (24.03) &0.243 (\textbf{0.231}) \\
		\hline
	\end{tabular}
	\label{tab1}
\end{table}
The results are presented in Table \ref{tab1}. In this table, T1\_test represents the short-utterance SLR task, and T2\_dev and T\_test represent the cross-channel SLR task. Values in parentheses are the results of the LR backend-based model. These results show that the mismatch between the training and test sets in Task 2 was much more severe than in Task 1. These results confirm that the performances of the two baseline systems are comparable. Therefore, all adaptation models in the subsequent experiments were based on a neural-network-backend-model baseline system.

\subsubsection{Adaptation performance}
After adaptation training, we expect the distribution mismatch between the training and test to be reduced; hence, improved performance can be achieved. The performance of our proposed adaptation algorithm and several comparison algorithms are shown in Table \ref{tabcompareT1} for the test set of Task 1, and in Tables \ref{tabcompareT2dev} and \ref{tabcompareT2} for the dev and test sets of Task 2, respectively. In these tables, `D-MMD' and `D-CMD' are UDA algorithms based on direct minimizing latent feature distribution discrepancy between domains via MMD and CMD-based discrepancy measure according to Eqs. (\ref{eq_DMMD1}) and (\ref{eq_DMMD2}). `DANN` is based on domain adversarial neural network according to Eqs. (\ref{eq_adv})-(\ref{eq_minmax}). `WDGRL' is also an adversarial UDA; however, based on Wasserstein distance-guided representation learning according to Eqs. (\ref{eqwdapp})-(\ref{eqwdgrl}). `NOT' and `NPOT' are our proposed neural optimal transport and neural partial optimal transport based UDA algorithms as introduced in Section \ref{sect_POT}.       
\begin{table}[tb]
	\centering
	\caption{Performance on adaptation model on task1 test set.}
	\begin{tabular}{|c||c||c|}
		\hline
		Methods &EER(\%) &Cavg\\
		\hline
		\hline
		D-MMD  &5.233&0.0513\\
		\hline
		D-CMD  &5.528&0.0557\\
		\hline
		DANN  &5.300&0.0524\\
		\hline
		\hline
		\rowcolor[gray]{0.9}
		WDGRL  &5.083&0.0509\\
		\hline
		\rowcolor[gray]{0.8}
		NOT (ours)  &5.056&0.0507\\
		\hline
		\rowcolor[gray]{0.7}
		NPOT (ours) &5.011&0.0503\\
		\hline
	\end{tabular}
	\label{tabcompareT1}
\end{table}
\begin{table}[tb]
	\centering
	\caption{Performance on adaptation model on task2 dev set.}
	\begin{tabular}{|c||c||c|}
		\hline
		Methods &EER(\%) &Cavg\\
		\hline
		\hline
		D-MMD  &14.43&0.1439\\
		\hline
		D-CMD  &13.67&0.1326\\
		\hline
		DANN  &12.67&0.1137\\
		\hline
		\hline
		\rowcolor[gray]{0.9}
		WDGRL  &8.30&0.0846\\
		\hline
		\rowcolor[gray]{0.8}
		NOT (ours)  &7.567&0.0777\\
		\hline
		\rowcolor[gray]{0.7}
		NPOT (ours)  &5.267&0.0499\\
		\hline
	\end{tabular}
	\label{tabcompareT2dev}
\end{table}
\begin{table}[tb]
	\centering
	\caption{Performance on adaptation model on task2 test set.}
	\begin{tabular}{|c||c||c|}
		\hline
		Methods &EER(\%) &Cavg\\
		\hline
		\hline
		D-MMD  &13.41&0.1364\\
		\hline
		D-CMD  &13.39&0.1427\\
		\hline
		DANN  &14.37&0.1436\\
		\hline
		\hline
		\rowcolor[gray]{0.9}
		WDGRL  &9.296&0.0916\\
		\hline
		\rowcolor[gray]{0.8}
		NOT (ours)  &8.139&0.0828\\
		\hline
		\rowcolor[gray]{0.7}
		NPOT (ours)  &6.704&0.0684\\
		\hline
	\end{tabular}
	\label{tabcompareT2}
\end{table}

In different UDA algorithms, the weighting coefficients for controlling the balance between source classification loss and domain alignment loss are different, and they are empirically set $\lambda =5.0$ for D-MMD (and D-CMD), $\lambda=0.5$ for adversarial loss in DANN. In WDGRL, the gradient penalty was set to $\eta=20$, and the adversarial loss weight was set to $\lambda=0.1$. In the proposed neural-OT and neural partial-OT-based adaptation models, the parameter settings were $\lambda=1.0$, $\alpha=0.001$, $\beta=5$ and $\tau=1$. In all these UDA algorithms, the entropy loss for the target domain data defined in Eq. (\ref{eq_entropytarget}) were equally weighted, with a coefficient $0.05$. Comparing the results in Tables \ref{tab1}, \ref{tabcompareT1}, \ref{tabcompareT2dev} and \ref{tabcompareT2}, all the UDA algorithms significantly improve the performance. In addition, from the results of `WDGRL,' `NOT,' and `NPOT,' we can see that the performance with OT-based discrepancy measure showed the best performance (please note that in WDGRL, an OT-inspired domain critic loss function is adopted), and improved with a large margin on both Tasks 1 and 2 compared with conventional UDA algorithms. In the test set for Task 1, all the performances of `WDGRL,' `NOT,' and `NPOT' are comparable. However, there is a large improvement for both the dev and test sets of Task 2 and the improvement in adaptation in the cross-channel condition was much larger than that in Task 1.
\subsubsection{Visualization of language feature distributions}
\begin{figure}[tb]
	\centering
	\includegraphics[width=6cm, height=6cm]{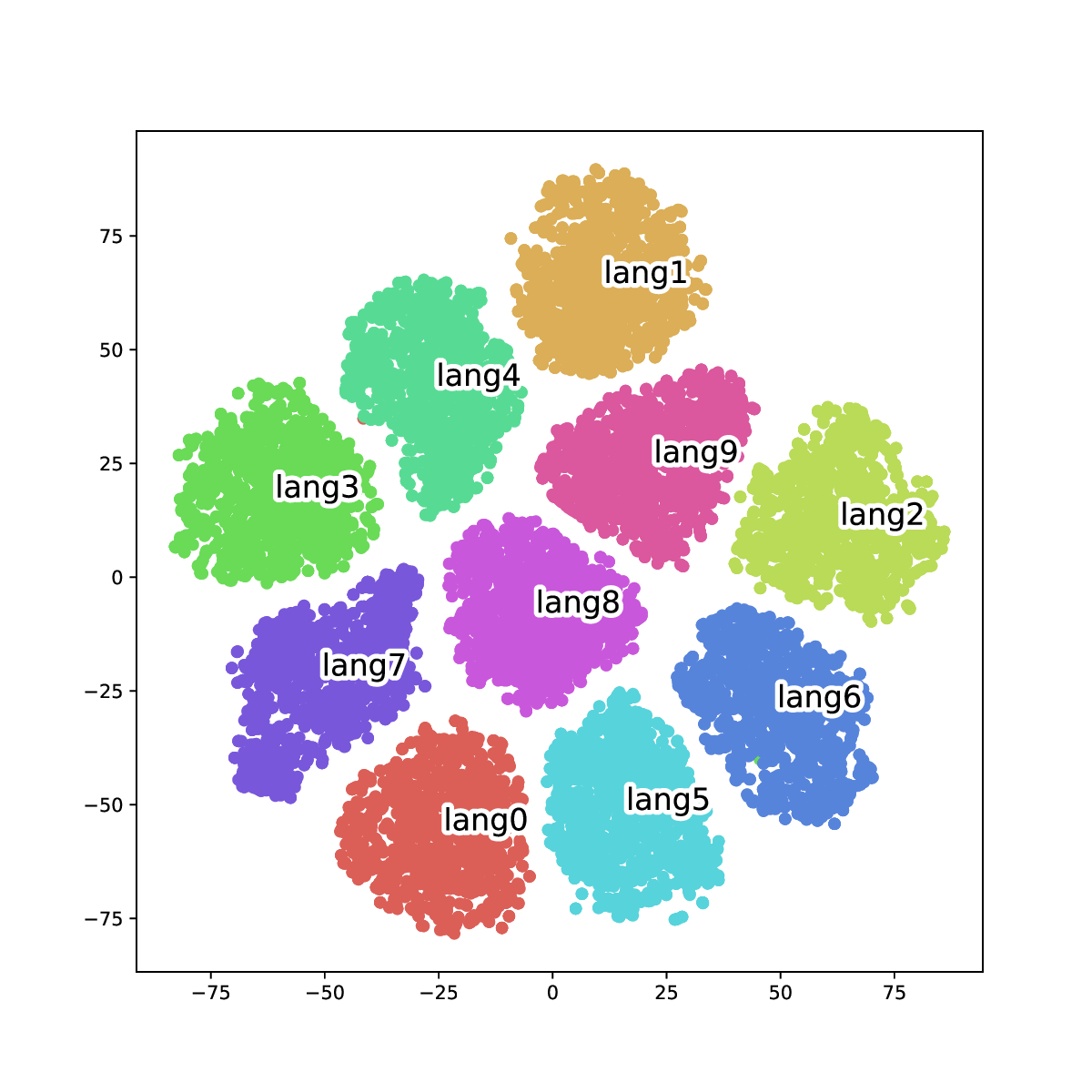}
	\caption{Language cluster distributions based on the t-SNE for training set}
	\label{figtrain}
\end{figure}

\begin{figure*}[tb]
	\centering
	\includegraphics[width=10cm, height=5cm]{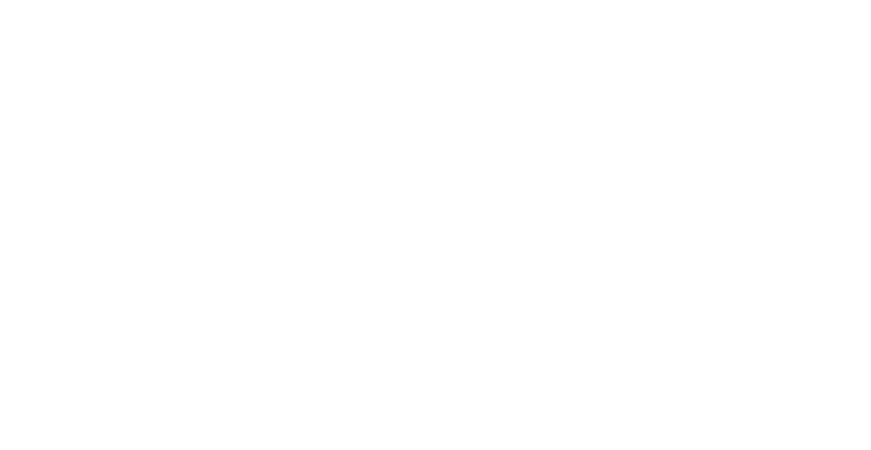}
	\caption{Language cluster distributions based on the t-SNE for test set task 1: (a) No adaptation, and (b) with adaptation.}
	\vspace{5mm}
	\label{figT1}
\end{figure*}

\begin{figure*}[tb]
	\centering
	\includegraphics[width=10cm, height=5cm]{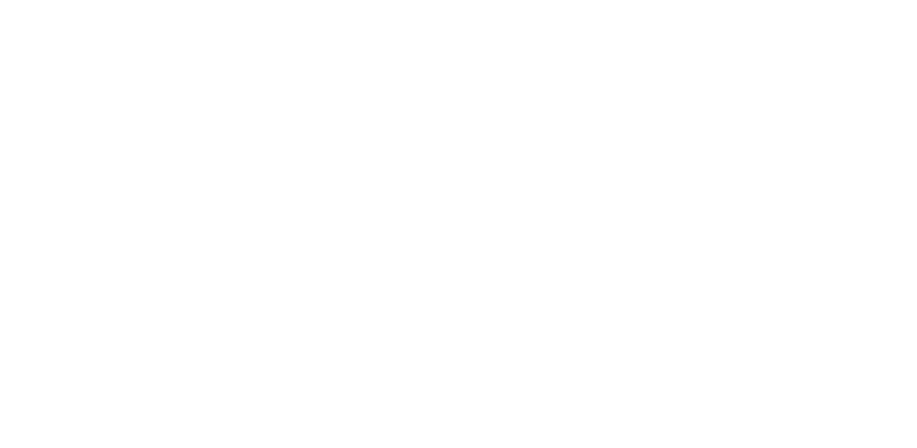}
	\caption{Language cluster distributions based on the t-SNE for test set task 2: (a) No adaptation, and (b) with adaptation.}
    \vspace{5mm}
	\label{figT2}
\end{figure*}

\begin{figure*}[tb]	
	\centering
	\includegraphics[width=12cm, height=4.5cm]{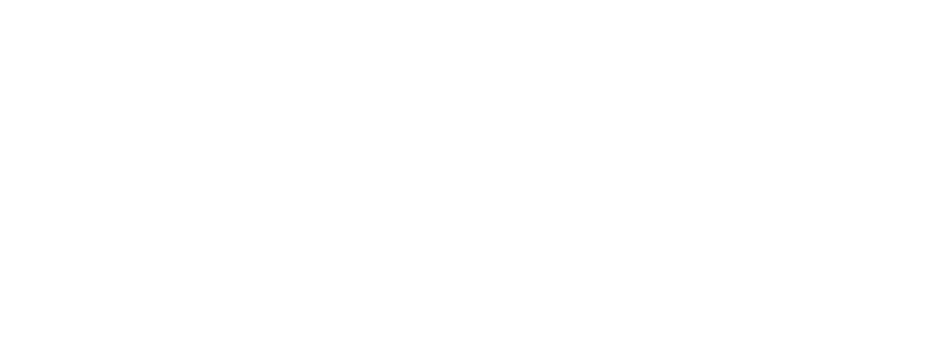}
	\caption{Language cluster distributions: (a) No adaptation, (b) adaptation based on NOT, and (c) based on partial NPOT. `lang1\_tr' and `lang1\_tt': Language 1 from training and test sets, respectively; `lang7\_tr' and `lang7\_tt': Language 7 from training and test sets, respectively.}
	\vspace{5mm}
	\label{fig_JBFrm}
\end{figure*}

We visually checked the language cluster distributions of the latent features (features input into the classifier layer, as shown in Fig. \ref{figConv}) for both training and test data sets on models before and after adaptation. This visualization is based on $ t $-distributed Stochastic Neighbor Embedding (t-SNE) \cite{Maaten2008}. A visualization of the feature distribution is shown in Figs. \ref{figtrain}, \ref{figT1}, \ref{figT2}. In these figures, clusters with different colors are distributions of samples (utterances) from languages corresponding to: `Kazakh': lang0, `Tibetan': lang1, `Uyghur': lang2, `Cantonese': lang3, `Indonesian': lang4, `Japanese': lang5, `Korean': lang6, `Russian': lang7, `Vietnamese': lang8, `Mandarin': lang9. From Fig. \ref{figtrain}, we note that speaker clusters are well separated based on the latent representation for the training data set. However, there were large overlaps between different language clusters for the test sets for Tasks 1 and 2, as depicted in Figs. \ref{figT1}-a and \ref{figT2}-a. In particular, the overlap was more significant in the test set of Task 2, mainly owing to the severe mismatch of the recording channels (Fig. \ref{figT2}-a). After model adaptation, the distribution of language clusters was well separated for the test sets of both Task 1 and 2 (comparing the cluster distributions in Figs. \ref{figT1} and \ref{figT2}). For a further clear view of the adaptation effect, the feature distributions of only two languages (with language ID labeled as ``lang1" and ``lang7") are depicted in Fig. \ref{fig_JBFrm}. In this figure, the labels with ``\_tr" and ``\_tt" denotes that samples are from training and test datasets (task 2), respectively. From Fig. \ref{fig_JBFrm}-a, we can note that there is a significant distribution gap between clusters belonging to the same language (pairs of lang7\_tr vs. lang7\_tt, and lang1\_tr vs. lang1\_tt). Because the classification model is trained with the training dataset, it is not surprising that the performance of the testing dataset is degraded. After adaptation training using our NOT algorithm, the distribution discrepancy between the training and test sets was reduced (as depicted in Fig. \ref{fig_JBFrm}-b). Furthermore, after adaptation training with our new NPOT algorithm, the clusters of the testing dataset of the same language were further pushed to overlap with the training dataset (as depicted in Fig. \ref{fig_JBFrm}-c). From the visualization, we can intuitively note that adaptation aids in reducing the intra-class feature distribution gap between domains while maintaining inter-class discrimination. In addition, from this visualization, we note that there are still language cluster overlaps. For example, in Fig. \ref{figT1}-b for the test set in Task 1, samples of `lang1' are not separated well from other language types; in Fig. \ref{figT2}-b, distributions of samples of `lang5' and `lang9' still have a significant overlap.            


\subsubsection{Effect of hyper-parameters in adaptation performance}
In the proposed adaptation algorithm, several hyperparameters play crucial roles in the adaptation performance. In Eq. (\ref{eq_softw}), $\beta$ and $\tau$ decide the shrinkage property of the sigmoid function for transport coupling or determine the relative importance of transport coupling in OT. As we investigated, there is a considerable range in selecting these hyperparameters that can maintain a relatively stable performance. In this study, we set their values as $\beta=5$ and $\tau=1$. In Eq. (\ref{eq_TLoss}), $\lambda$ controls the balance between the source domain classification loss and joint distribution alignment loss in adaptation learning. In Eq. (\ref{eq_jdist}), $\alpha$ determines the contribution of the label distribution difference between the source and target domains during cost estimation for OT optimization, and when $\alpha =0$, the model degenerates to a latent feature-based OT. Intuitively, when the prediction of the target label is accurate, a considerable $\alpha$ can be set. Hence, domain alignment by taking the label distribution difference in the cost function estimation (refer to Eq. (\ref{eq_jdist})) can aid in improving adaptation performance. Contrary, when there are several errors in the prediction of the target label, a smaller $\alpha$ is preferable; hence, the domain alignment is mainly determined by the latent feature distributions of the source and target datasets. As depicted in Fig. \ref{figcoupling}, hyper-parameters $\alpha$ and $\lambda$ are involved in a bi-level optimization process for the convenience of investigating the adaptation effect by considering the label distribution difference; we conducted experiments on the test set of Task 2 by varying $\alpha$ with fixed $\lambda=1$. The results are shown in table \ref{tab3}.          
\begin{table}[tb]
\centering
\caption{Effect of hyper-parameters on adaptation performance for test set of task 2.}
\begin{tabular}{|c||c||c|}
\hline
 Hyper-parameter $\alpha$ &EER\% &Cavg\\
\hline
0.0  &7.907&0.0801\\
\hline
1.0e-5  &7.074&0.0721\\
\hline
1.0e-4  &6.713&0.0667\\
\hline
5.0e-4  &\textbf{6.519}&\textbf{0.0663}\\
\hline
1.0e-3  &6.704&0.0684\\
\hline
5.0e-3  &6.972&0.0689\\
\hline
1.0e-2  &9.000&0.0895\\
\hline
5.0e-2  &9.324&0.0954\\
\hline
1.0e-1  &11.12&0.1134\\
\hline
\end{tabular}
\label{tab3}
\end{table}
We note from this table that the latent feature distribution adaptation is more important and effective than the label distribution adaptation. Even without label distribution adaptation, the performance was satisfactory. However, adding label distribution information with a small weight to the domain alignment improves the performance.   
\subsubsection{Optimal transport couple weighting}
\begin{figure}[tb]
	\vspace{5mm}
	\centering
	\includegraphics[width=7cm, height=4.5cm]{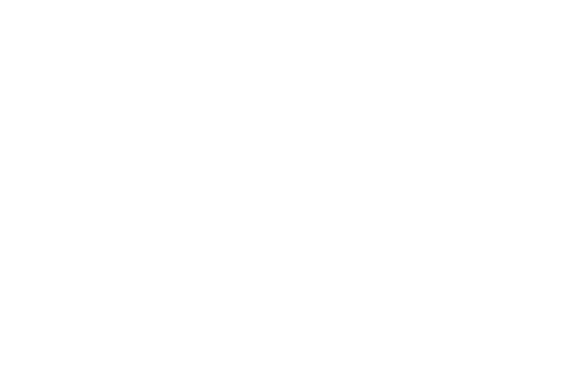}
	\caption{Optimal transportation couple weighting in adaptation (a) for task1, and (b) for task2.}
	\label{figWeight}	
\end{figure}

After the adaptation model was trained, we selected 64 samples from a mini-batch and sorted them according to their labels (in the adaptation learning stage, the target labels were unknown). The label space of the training data comprised 10 classes. The test set in Task 1 used the same label space as in the training data. However, for the test set in Task 2, the label space included only six classes. The partial coupling weights for the two tasks are depicted in Fig. \ref{figWeight}. 
In this figure, \{C0-C9\} denotes the 10 language labels. For Task 2, only partial labels (\{C1, C2, C5, C7, C8, C9\}) were used in the test set. This figure depicts well-structured couplings between samples from the same class, that is, the transports between samples with high coupling weights are performed during model adaptation learning. However, from this figure, we observe that a few couplings with mistakes tried to adapt to the wrong categories. For example, in Task 1, sample from class C1 were not transported well in adaptation training, and in Task 2, samples from class C5 were incorrectly coupled during OT. Regarding Figs. \ref{figT1} and \ref{figT2}, we can observe that the incorrect couplings are extremely discriminating classes with high confusion with other language categories.   

\section{Discussion and conclusion}
\label{sect_conclude}
In this study, we propose a novel unsupervised neural adaptation model for SLR. The X vector in this model is extracted from a well-trained deep neural network for language recognition based on the cross-entropy criterion. Following the basic domain adaptation theory, a joint distribution adaptation framework is proposed for both latent features and label distributions. We adopted an OT-based distribution discrepancy measure inspired by joint distribution optimal transport studies to design a neural domain alignment loss besides the classification loss for source domain data. Moreover, considering that the labeling space may differ from the training dataset, we designed a weighting scheme on the optimal transport-coupling matrix to fulfill the function of partial optimal transport in the domain alignment. The experimental results demonstrated that adaptation reduced the distribution difference between the source and target datasets and SLR performance achieved significant improvement, particularly on cross-channel SLR tasks. This study confirms that information on both latent features and label distributions should be combined in a domain alignment.  

The proposed UDA method has some limitations. As depicted in Figs. \ref{figT1}-b and \ref{figT2}-b, we can observe that there are overlaps in cluster distributions between different languages even after the adaptation process. Additionally, from Fig. \ref{figWeight}, we note that the existing negative transports result in negative adaptation. In current UDA methods, because the target labels are unknown, it is difficult to guarantee that the optimal coupling is class-wise in OT, and the discriminative information of some classes is inevitably degraded, although the domain discrepancy is reduced. More specifically, there is a trade-off between domain discrepancy reduction and increasing the discriminative power between classes. Another limitation of our study is that we used a fixed weighting coefficient for the domain alignment loss (refer to (Eq. \ref{eq_TLoss})). In future work, we will adaptively adjust the alignment loss to control the trade-off between domain adaptation and class discrimination.



\begin{thebibliography}{1}

\bibitem{Mabin2013}
H. Li, B. Ma, B., K. Lee, ``Spoken Language Recognition: From Fundamentals to Practice," \emph{Proceedings of the IEEE}, 101 (5), pp. 1136-1159, 2013.

\bibitem{Mabin2007c}
H. Li, B. Ma, B. C. Lee, ``A Vector Space Modeling Approach to Spoken Language Identification," \emph{IEEE Transactions on Audio, Speech and Language Processing}, 15 (1), pp. 271-284, 2007.

\bibitem{Mabin2007b}
B. Ma, H. Li, R. Tong, ``Spoken Language Recognition with Ensemble Classifiers," \emph{IEEE Transactions on Audio, Speech and Language Processing}, 15 (7), pp. 2053-2062, 2007.

\bibitem{Lee2016IS}
K. Lee, H. Li, L. Deng, et al., ``The 2015 NIST Language Recognition Evaluation: the Shared View of I2R, Fantastic4 and SingaMS," in \emph{Proc. of INTERSPEECH}, pp. 3211-3215, 2015.

\bibitem{Dehak2011}
N. Dehak, P. Kenny, R. Dehak, P. Dumouchel, and P. Ouellet, ``Front-end factor analysis for speaker verification," \emph{IEEE Transactions on Audio, Speech, and Language Processing}, vol. 19, no. 4, pp. 788-798, 2011.

\bibitem{Prince2007}
S. Prince and J. Elder, ``Probabilistic linear discriminant analysis for inferences about identity," in \emph{Proc. of IEEE International Conference on Computer Vision (ICCV)}, pp. 1-8, 2007.

\bibitem{Variani2014}
E. Variani, X. Lei, E. McDermott, I. L. Moreno and J. Gonzalez-Dominguez, ``Deep neural networks for small footprint text-dependent speaker verification," in \emph{Proc. of ICASSP},  pp. 4052-4056, 2014.

\bibitem{Snyder2018}
D. Snyder, D. Garcia-Romero, G. Sell, D. Povey, and S. Khudanpur, ``X-vectors: Robust dnn embeddings for speaker recognition," in \emph{Proc. of ICASSP}, pp. 5329-5333, 2018.

\bibitem{RichardsonIS}
R. Richardson, D. Reynolds, N. Dehak, ``A Unified Deep Neural Network for Speaker and Language Recognition," in \emph{Proc. of INTERSPEECH}, pp. 1146-1150, 2015.

\bibitem{RichardsonIEEE}
R. Richardson, D. Reynolds, N. Dehak, ``Deep Neural Network Approaches to Speaker and Language Recognition," \emph{IEEE Signal Processing Letters}, 22 (10), pp. 1671-1675, 2015.

\bibitem{Ranjan2016}
S. Ranjan, C. Yu, C. Zhang, F. Kelly and J. Hansen, ``Language recognition using deep neural networks with very limited training data," in \emph{Proc. of ICASSP}, pp. 5830-5834, 2016.

\bibitem{Moreno2016}
I. Lopez-Moreno, J. Gonzalez-Dominguez, D. Martinez, O. Plchot, J. Gonzalez-Rodriguez, P. J. Moreno, ``On the use of deep feedforward neural networks for automatic language identification," \emph{Computer Speech \& Language}, Vol.40, pp. 46-59, 2016.

\bibitem{Moreno2014}
I. Lopez-Moreno, J. Gonzalez-Dominguez, O. Plchot, D. Martinez, J. Gonzalez-Rodriguez and P. Moreno, "Automatic language identification using deep neural networks," in \emph{Proc. of ICASSP}, pp. 5374-5378, 2014.

\bibitem{Diez2015}
A. Lozano-Diez, R. Zazo Candil, J. G. Dominguez, D. T. Toledano and J. G. Rodriguez, ``An end-to-end approach to language identification in short utterances using convolutional neural networks," in \emph{Proc. of INTERSPEECH}, pp. 403-407, 2015.

\bibitem{Fernando2017}
 S. Fernando, V. Sethu, E. Ambikairajah and J. Epps, ``Bidirectional Modelling for Short Duration Language Identification," in \emph{Proc. of INTERSPEECH}, pp. 2809-2813, 2017.

\bibitem{Geng2016}
W. Geng, W. Wang, Y. Zhao, X. Cai and B. Xu, ``End-to-End Language Identification Using Attention-Based Recurrent Neural Networks," in \emph{Proc. of INTERSPEECH}, pp. 2944-2948, 2016.

\bibitem{Romero2014ODS}
D. G. Romero, A. McCree, S. Shum, N. Brummer, and C. Vaquero, ``Unsupervised domain adaptation for i-vector speaker recognition," in \emph{Proc. of the Speaker and Language Recognition Workshop (Odyssey)}, pp. 260-264, 2014.

\bibitem{Romero2014SLT}
D. G. Romero, X. Zhang, A. McCree, and D. Povey, ``Improving speaker recognition performance in the domain adaptation challenge using deep neural networks," in \emph{Proc. of IEEE Spoken Language Technology Workshop (SLT)}, pp. 378-383, 2014.

\bibitem{LeeICASSP2019}
K. Lee, Q. Wang, and T. Koshinaka, ``The coral+ algorithm for unsupervised domain adaptation of plda," in \emph{Proc. of ICASSP}, pp. 5821-5825, 2019.

\bibitem{Bousquet2019}
P. Bousquet, M. Rouvier, ``On robustness of unsupervised domain adaptation for speaker recognition," in \emph{Proc. of INTERSPEECH}, pp. 2958-2962, 2019.

\bibitem{MMD}
B. Scholkopf, J. Platt, T. Thomas, ``A Kernel Method for the Two-Sample-Problem," in \emph{Proc. of the International Conference on Neural Information Processing Systems (NIPS)}, vol. 19, pp. 513-520, 2006.

\bibitem{CMD}
W. Zellinger, T. Grubinger, E. Lughofer, T. Natschlager, and S. Saminger-Platz, ``Central moment discrepancy (cmd) for domain-invariant representation learning," \emph{arXiv preprint arXiv:1702.08811}, 2017.

\bibitem{DANN2016}
Y. Ganin, E. Ustinova, H. Ajakan, P. Germain, H. Larochelle, F. Laviolette, M. Marchand, and V. Lempitsky, ``Domain-adversarial training of neural networks," \emph{Journal of Machine Learning Research},  vol. 17, no. 1, pp. 1-35, 2016.

\bibitem{Badr2020}
B. Abdullah, T. Avgustinova, B. Mobius, D. Klakow, ``Cross-Domain Adaptation of Spoken Language Identification for Related Languages: The Curious Case of Slavic Languages," in \emph{Proc. of INTERSPEECH}, pp. 477-481, 2020.

\bibitem{Goodfellow2014}
I. Goodfellow, J. Pouget-Abadie, M. Mirza, B. Xu, D. Warde-Farley, S. Ozair, A. Courville and Y. Bengio, ``Generative adversarial nets," in \emph{Proc. the International Conference on Neural Information Processing Systems (NIPS)}, pp. 2672-2680, 2014.

\bibitem{Shen2018}
J. Shen, Y. Qu, W. Zhang, and Y. Yu, ``Wasserstein distance guided representation learning for domain adaptation," in \emph{Proc. of the Thirty-Second AAAI Conference on Artificial Intelligence (AAAI-18)}, no. 497, pp. 4058-4065, 2018.

\bibitem{CourtyIEEE2017}
N. Courty, R. Flamary, D. Tuia, A. Rakotomamonjy, ``Optimal transport for domain adaptation," \emph{IEEE Transactions on Pattern Analysis and Machine Intelligence}, 39 (9), pp. 1853-1865, 2017.

\bibitem{Peyre2018}
G. Peyre, M. Cuturi, ``Computational Optimal Transport," \emph{ArXiv:1803.00567}, 2018.

\bibitem{CourtyNIPS2017}
N. Courty, R.  Flamary, A. Habrard, A. Rakotomamonjy, ``Joint distribution optimal transportation for domain adaptation," in \emph{Proc. of the International Conference on Neural Information Processing Systems (NIPS)}, pp. 3733-3742, 2017.

\bibitem{DamodaranECCV2018}
B. Damodaran, B. Kellenberger, R. Flamary, D. Tuia, and N. Courty, ``DeepJDOT: Deep joint distribution optimal transport for unsupervised domain adaptation," in \emph{Proc. of the European Conference on Computer Vision (ECCV)}, pp. 447-463, 2018.

\bibitem{Lin2021}
H. Lin, H. Tseng, X. Lu, Y. Tsao, ``Unsupervised Noise Adaptive Speech Enhancement by Discriminator-Constrained Optimal Transport," in \emph{Proc. of the International Conference on Neural Information Processing Systems (NeurIPS)}, pp. 19935-19946, 2021.

\bibitem{LuICASSP2021}
X. Lu, P. Shen, Y. Tsao, H. Kawai, ``Unsupervised Neural Adaptation Model Based on Optimal Transport for Spoken Language Identification," in \emph{Proc. of ICASSP}, pp. 7213-7217, 2021.

\bibitem{OLR2020}
http://index.cslt.org/mediawiki/index.php/OLR\_Challenge\_2020

\bibitem{Chuang2020}
C. Chuang, A. Torralba, and S. Jegelka,``Estimating generalization under distribution shifts via domain-invariant representations," in \emph{Proc. of International
Conference on Machine Learning (ICML)}, pp. 1984-1994, 2020.

\bibitem{LiuBound}
X. Liu, et al., ``Adversarial Unsupervised Domain Adaptation with Conditional and Label Shift: Infer, Align and Iterate," in \emph{Proc. of IEEE/CVF International Conference on Computer Vision (ICCV)}, pp. 10347-10356, 2021.

\bibitem{Kouw2019}
W. M. Kouw and M. Loog, ``A Review of Domain Adaptation without Target Labels," \emph{IEEE transactions on pattern analysis and machine intelligence}, 43(3), pp. 766-785, 2019.

\bibitem{PanMMD2009}
S. Pan, I. Tsang, J. Kwok and Q. Yang, ``Domain Adaptation via Transfer Component Analysis," in \emph{IEEE Transactions on Neural Networks}, vol. 22, no. 2, pp. 199-210, Feb. 2011.

\bibitem{LongMMD2013}
M. Long, J. Wang, G. Ding, J. Sun, and P. S. Yu, ``Transfer feature learning with joint distribution adaptation," in \emph{Proc. of the IEEE International Conference on Computer Vision}, pp. 2200-2207, 2013.

\bibitem{WGAN2017}
M. Arjovsky, S. Chintala, and L. Bottou, ``Wasserstein generative adversarial networks," in \emph{Proc. of International Conference on Machine Learning}, pp. 214-223, 2017.

\bibitem{VillanoBook}
C. Villani, Optimal transport: old and new, volume 338. Springer, 2009

\bibitem{ImprovedGAN}
I. Gulrajani, F. Ahmed, M. Arjovsky, V. Dumoulin, and A. Courville, ``Improved training of wasserstein gans," \emph{arXiv preprint arXiv: 1704.00028}, 2017.

\bibitem{Figalli2010}
A. Figalli, ``The Optimal Partial Transport Problem," \emph{Arch Rational Mech Anal}, 195, pp. 533-560, 2010.

\bibitem{Chapel2020}
L. Chapel, M. Alaya, and G. Gasso, ``Partial optimal transport with applications on positive unlabeled learning," in \emph{Proc. of the International Conference on Neural Information Processing Systems (NeurIPS)}, vol. 33, pp. 2903-2913, 2020.

\bibitem{SSLEntropy}
Y. Grandvalet, and Y. Bengio, ``Semi-supervised learning by entropy minimization," in \emph{Proc. of International Conference on Neural Information Processing Systems (NIPS)}, pp. 529-536, 2004.

\bibitem{WD2019}
Z. Tang, D. Wang, L. Song, ``AP19-OLR Challenge: Three Tasks and Their Baselines," in \emph{Proc. of Asia-Pacific Signal and Information Processing Association Annual Summit and Conference (APSIPA ASC)}, pp. 1917-1921, 2019

\bibitem{WD2020}
Z. Li, M. Zhao, Q. Hong, L. Li, Z. Tang, D. Wang, L. Song, C. Yang, ``AP20-OLR Challenge: Three Tasks and Their Baselines," \emph{CoRR abs/2006.03473}, 2020.


\bibitem{Adam}
Diederik P. Kingma, Jimmy Ba, ``Adam: A Method for Stochastic Optimization," \emph{the 3rd International Conference on Learning Representations (ICLR)}, 2015.

\bibitem{Maaten2008}
L. Maaten, G. Hinton, ``Visualizing Data Using t-SNE," \emph{Journal of Machine Learning Research}, 9 (86), pp. 2579-2605, 2008.
\end{thebibliography}

\end{document}